\documentclass[namedreferences]{solarphysics}
\usepackage[optionalrh]{spr-sola-addons} 
\usepackage{graphicx}        
\usepackage{color}           
\usepackage{url}             

\newcommand{\etal}{{\it et al.}}

\newcommand{\aap}{    {\it Astron. Astrophys.}}

\newcommand{\apj}{    {\it Astrophys. J.}}
\newcommand{\apjl}{   {\it Astrophys. J. Lett.}}

\newcommand{\solphys}{{\it Solar Phys.}}

\begin{document}

\begin{article}

\begin{opening}

\title{Study of flare energy release using events with numerous type
III-like bursts in microwaves\\ {\it Solar Physics}}

\author{N.S.~\surname{Meshalkina}$^{1}$\sep
        A.T.~\surname{Altyntsev}$^{1}$\sep
        D.A.~\surname{Zhdanov}$^{1}$\sep
        S.V.~\surname{Lesovoi}$^{1}$\sep
         A.A.~\surname{Kochanov}$^{1}$\sep
         Y.H. ~\surname{Yan}$^{2}$\sep
         C.M. ~\surname{Tan}$^{2}$\sep
       }
\runningauthor{Altyntsev et al.} \runningtitle{Type III-like
bursts}

   \institute{$^{1}$Institute of Solar-Terrestrial Physics SB RAS,Lermontov St.\ 126A, Irkutsk 664033, Russia email: \url{nata@iszf.irk.ru} \\
       $^{2}$  Key Laboratory of Solar Activity, National Astronomical Observatories, CAS, Beijing 100012, China e-mail:
       \url{yyh@bao.ac.cn} \\
             }

\begin{abstract}
The analysis of narrowband drifting of type III-like structures in
radio bursts dynamic spectra allows to obtain unique information
about primary energy release mechanisms in solar flares. The SSRT
spatially resolved images and a high spectral and temporal
resolution allow direct determination not only the positions of
its sources but also the exciter velocities along the flare loop.
Practically, such measurements are possible during some special
time intervals when the SSRT (about 5.7~GHz) is observing the
flare region in two high-order fringes; thus, two 1D scans are
recorded simultaneously at two frequency bands. The analysis of
type III-like bursts recorded during the flare 14 Apr 2002 is
presented. Using-muliwavelength radio observations recorded by
SSRT, SBRS, NoRP, RSTN we study an event with series of several
tens of drifting microwave pulses with drift rates in the range
from -7 to 13 GHz s$^{-1}$. The sources of the fast-drifting
bursts were located near the top of the flare loop in a volume of
a few Mm in size. The slow drift of the exciters along the flare
loop suggests a high pitch-anisotropy of the emitting electrons.

\end{abstract}

\end{opening}


\section{Introduction}
     \label{S-Introduction}

During solar flares a large amount of stored magnetic field energy
is suddenly released, and transformed into heat, mass motions,
enhanced emission of electromagnetic radiation, and enhanced
fluxes of energetic particles. Radio emission reveals the presence
of nonthermal electrons, both in the acceleration region and along
their propagation path. In a recent study by
\inlinecite{Fleishman11} these two mentioned above components were
identified in the observational data and carefully separated.
Major results on the propagation of electron beams have been
obtained studying drifting bursts of the so-called type III, which
are characterized as narrow-band bursts whose frequency either
rapidly increases or drops with time. It is generally accepted
that the emission exciter frequency of type III bursts is
determined by the local plasma frequency, which is proportional to
the square root of the plasma density. Furthermore, a drift with
decreasing frequency is explained by the propagation of the
electron beams from the solar surface toward decreasing plasma
density, and an increase in frequency with time will correspond to
a downward movement (\opencite{Robinson};
\opencite{Aschwanden02}).

It is assumed that structures of subsecond duration ($<$ 1 s) in
radio emission are indicators of primary energy release processes
in solar flares. One of the probable mechanisms of the generation
of observable individual pulses in trains is the reconnection
process (\opencite{Kliem89}, \citeyear{Kliem95};
\opencite{Kliem00}; \opencite{Machado93}; \opencite{Aschwanden93},
\citeyear{Aschwanden04}). Radio bursts with fine temporal
structure and their response in HXR and radio emission have been
studied for more than 20 years (\opencite{Aschwanden92};
\opencite{Aschwanden98}; \opencite{Huang1999}; \opencite{Benz06}).
Although, as a rule, there is no unambiguous correlation between
the microwave profiles and the HXR pulses, on the subsecond
time-scale of subsecond pulses (SSP) agrees with HXR intensity
increases. Although SSPs are almost always associated with HXR
emission, the reverse is not true. HXR bursts are accompanied by
subsecond pulses very rare. This suggests the existence of some
special boundary conditions for a temporal fine structure to be
generated and for the emission to escape from the generation site.

Observations of subsecond pulses characterised by narrow-band
coherent emission provide a promising way to estimate the plasma
parameters at acceleration sites. Using one-dimensional brightness
distribution recorded with the \textit{Siberian Solar Radio
Telescope} (SSRT) interferometer in adjacent high-order fringes we
can measure the exciter displacement at two different frequencies
near 5.7 GHz in direction perpendicular the interference fringe
pattern.  Thus, it is possible to measure the exciter velocity and
plasma density gradient along one direction in the image plane.

Characteristics of drifting bursts in microwave range considerably
differ from the well-studied metric bursts of type III. At high
plasma densities a key question whether emission is detectable
despite the strong free-free absorption in the dense plasma
surrounding the fast-drifting exciter. Emission at the plasma
frequency cannot escape from the source. Weaker absorption at the
doubled plasma frequency can be estimated (\opencite{Dulk1985};
\opencite{Benz93}) as $\exp^{-\tau}$, where:

\begin{eqnarray}
 \tau=0.12\left(\frac{f}{1 MHz}\right)^2\left(\frac{T}{1
 MK}\right)^{-3/2}\left(\frac{H}{1 Mm}\right)
\end{eqnarray}

\noindent where $H$ is the transverse scale of variation in
density, $T$ is plasma temperature in the emission source, $f$ is
 frequency of the observed radio emission.

The SSRT observations with spatial resolution confirm that the
observed emission frequency of the fast-drifting bursts is close
to the doubled plasma frequency determined by the sources' plasma
density (\opencite{Meshalkina04}). For emission at the SSRT
working frequency (5.7 GHz) the exciter plasma density should
exceed $10^{11} \mathrm{cm}^{-3}$ and density scale height $H$
should be less than thousand km. This width is lower than the
flare loop's cross-sections observed in UV and X-rays. A favorable
condition to observe the coherent microwave emission is realized
when the source is located at large height near the top of the
dense hot loop so that the electromagnetic waves can escape across
the flare loop. Really, in the majority of events with the
subsecond pulses the sources have been located near loop tops
(\opencite{Meshalkina04}).

At high frequencies the positive drifts are predominant. This
direction of a frequency drift can be naturally explained by the
exciter moving from a loop top to denser plasma in the legs of the
loop. Observations of a specific type of the fast-drifting bursts,
so called U-bursts, shows, that the apparent frequency drifts can
be caused not only by an exciter moving through a density
gradient, but by a change of density in a steady emission source
(\opencite{Altyntsev2003a}). So, to interpret fast-drifting bursts
in the microwaves we should distinguish between effects of
temporal and spatial changes of plasma density. In this context
the analysis of the events with a large number of fast-drifting
bursts is promising.

Note that spatially-resolved trains of short drifting bursts are
very rarely recorded simultaneously with their dynamic spectra:
until now, the only such event was found on 30 March 2001
\cite{Altyntsev2007}. For a half-minute interval, the drift rates
of 67 drifting subsecond pulses were $-$10 to 20 GHz s$^{-1}$ and
widely scattered around an average value of 6 GHz s$^{-1}$. The
estimated exciter velocities from the drift rates, without
correcting for this mean, results were in a wide range, some even
exceeding the speed of light. To solve this discrepancy, it was
suggested that the overall drift was due to increase of density at
the emission site. Plasma influx can result from a magnetic
reconnection process. For the one-dimensional approximation
(Sweet-Parker model) the current-sheet thickness should be of
order 20 m and the reconnection rate $\alpha\approx 3.6 \times
10^{-5}$ of the Alfv\'{e}n velocity.

Recently, we discovered a second event (14 Apr 2002) with a large
number of drifting bursts recorded simultaneously with the SSRT
interferometer (at two frequencies) and the Chinese
spectropolarimeter. The presence of a large number of drifting
bursts gives us the chance to extract the component of frequency
drift reflecting a plasma density increase in the emission region.
The purpose of this study is to analyze the 14 Apr 2002 event and
compare the results for the two events as well as estimating the
plasma parameters in the fast drifting burst sources.

\section{Instruments}

The dynamic spectra of the microwave burst were recorded with the
\textit{ Solar Broadband Radio Spectrometer} (SBRS, 5.2 -- 7.6
GHz) at the Huairou Solar Observing Station of the National
Astronomical Observatories of China \cite{Fu95}. The
single-channel bandwidth of the SBRS is 20 MHz, and the temporal
resolution is 5 ms \cite{Ji03}. We used the SBRS total flux
records to extract individual bursts.

The spatial structures were recorded with the SSRT
\cite{Grechnev03}. The SSRT consists of two linear antenna arrays
-- East - West (EW) and North - South (NS) -- and operates in the
5.67 -- 5.79 GHz range. The right -- (RCP) and left-handed (LCP)
circularly polarized components are recorded alternately, 7 ms
each. The SSRT produces two-dimensional full-disk images every 3
-- 5 min, and observes microwave bursts in the one-dimensional
(1D) mode with the EW and NS linear interferometers independently.
The 1D mode has a temporal resolution of 14 ms. The time profiles
of radio flux variation in the intensity (I=R+L) and circular
polarization (V=R-L) channels are posted on a regular basis, at
the Institute of Solar-Terrestrial Physics Radio Astronomical
Observatory website\footnote{http://www.ssrt.org.ru/ and
http://ssrt.iszf.irk.ru/fast/}.

The SSRT receiver system is a 120-MHz-band spectrum analyzer
implemented as an acousto-optic receiver. It has 250 frequency
channels, which correspond to fans of knife edge beams for the NS
and EW arrays. The bandwidth of a single frequency channel is 0.48
MHz. The response at each frequency corresponds to emission from a
narrow strip on the solar disk, whose position and width depend on
the frequency. During the event under study, the NS beam was 24.3
arc sec, the EW beam 15.3 arc sec wide.

    The microwave data were taken from the \textit{Nobeyama Radio Polarimeters}
(NoRP, \opencite{Torii}; \opencite{Shibasaki1979};
\opencite{Nakajima85}). The NoRP measures the fluxes at 1, 2,
3.75, 9.4, 17, 35, 80 GHz with a temporal resolution of 1 s for
steady mode and 0.1 s for flare mode. The Nobeyama Radioheliograph
(NORH, \opencite{Nakajima94}) observed the flare at 17 and 34 GHz.
The cadence of the data used here is 0.1 s. The NoRH beam sizes
were 11 $\times$ 17$^{\prime\prime}$ at 17 GHz and 8 $\times$
11$^{\prime\prime}$ at 34 GHz at the time of the flare.

We also used the \textit{Radio Solar Telescope Network} (RSTN)
data. One-second temporal resolution data were taken at eight
frequencies (0.245, 0.41,0.61, 0.14, 0.27, 0.5, 0.88, 15.4 GHz)
from Learmonth station, Australia. We used GOES data and extreme
ultraviolet (EUV) observations of the lower corona by the EIT (195
\AA) \cite{Delaboudinière} onboard the SOHO. SOHO/MDI
\cite{Scherrer95} magnetograms represent photospheric magnetic
fields.

\section{Data analysis}

The SSRT can simultaneously register a flare region in two
interference orders at different frequencies of the SSRT band
turning on the observing direction to the baselines
\cite{Grechnev03}. The NS interferometer observed the 14 April
2002 flare region in two high-order fringes; thus, two 1D scans
were recorded simultaneously within two frequency intervals
centered at about 5.67 and 5.76 GHz. The EW linear interferometer
recorded the 1D data at 5.69 GHz. The methods to analyze 1D SSRT
scans have been described by \inlinecite{Altyntsev96} and
Altyntsev et al. (\citeyear{Altyntsev2003a},
\citeyear{Altyntsev2007}). The apparent sizes of pulse sources are
exceed the real sizes and are determined by the SSRT beam size and
by wave scattering in the low corona (\opencite{Altyntsev96};
\opencite{Meshalkina05}). The accuracy in measuring a relative
displacement of a source recorded simultaneously at two different
frequencies is estimated to be within 2 arcsec. This accuracy is
achieved by measuring their centroids, and it was verified by
modeling of the response with the observed noise level.

We have extracted 56 pulses from the dynamic spectrum of SBRS
crossing the SSRT band (5.67 -- 5.79 GHz) during a time interval
of 30 sec (05:36:26 -- 05:36:56). For each pulse we estimated
displacements relative to the background burst.

    \begin{figure}    
  \centerline{\includegraphics[width=\textwidth]{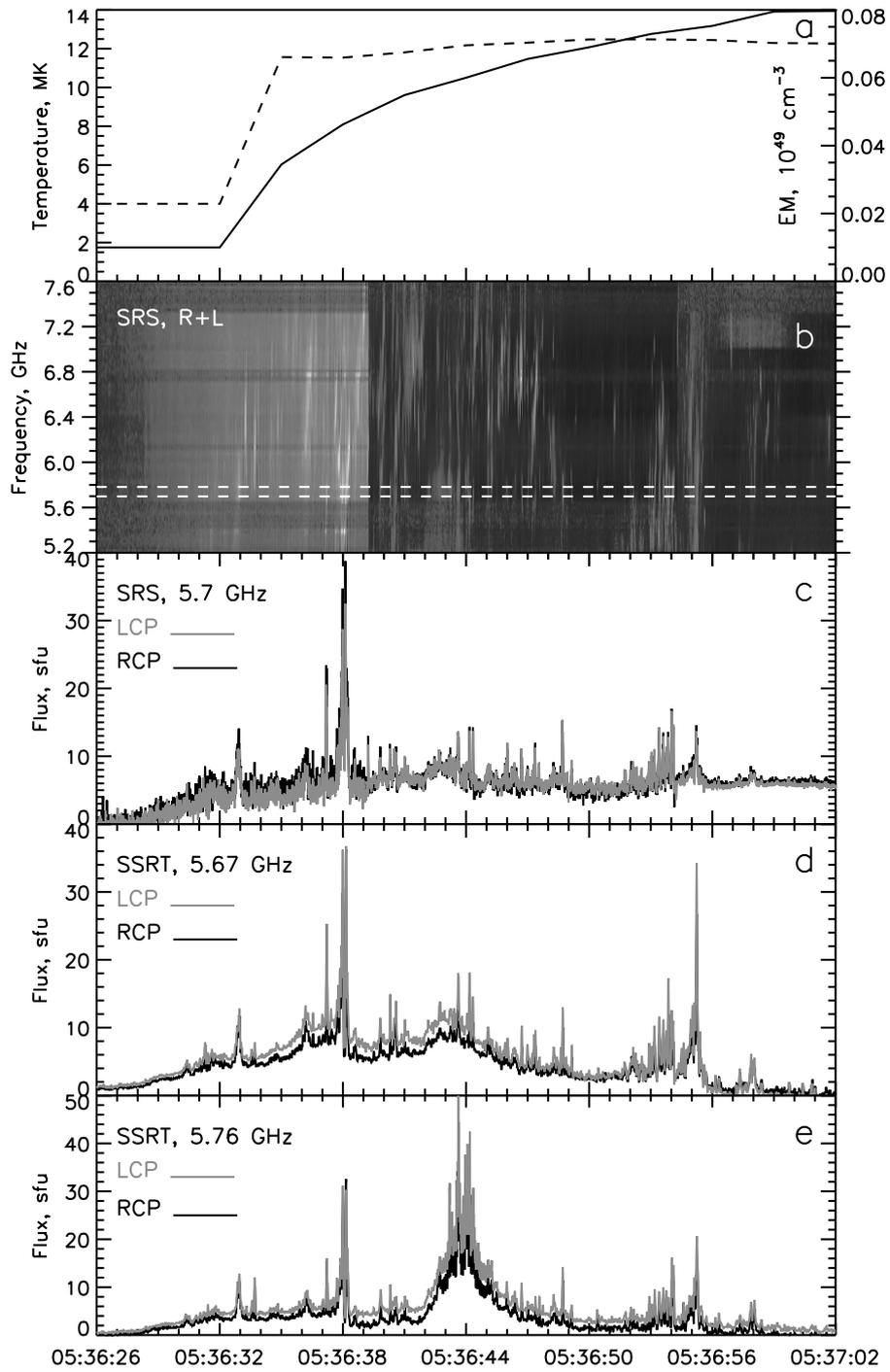}
              }
                \caption{a) temperature (dashed line) and emission measure (solid line)(GOES10 data);
                b) dynamic spectrum in intensity (5.2 - 7.6 GHz, SBRS);
                c) RCP and LCP time profiles from the spectropolarimeter
at the central frequency of SSRT band; (d), (e) RCP and LCP time
profiles at two frequencies recorded with SSRT.}
   \label{F1-simple}
   \end{figure}

\begin{figure}
     \centerline{\includegraphics[width=\textwidth]{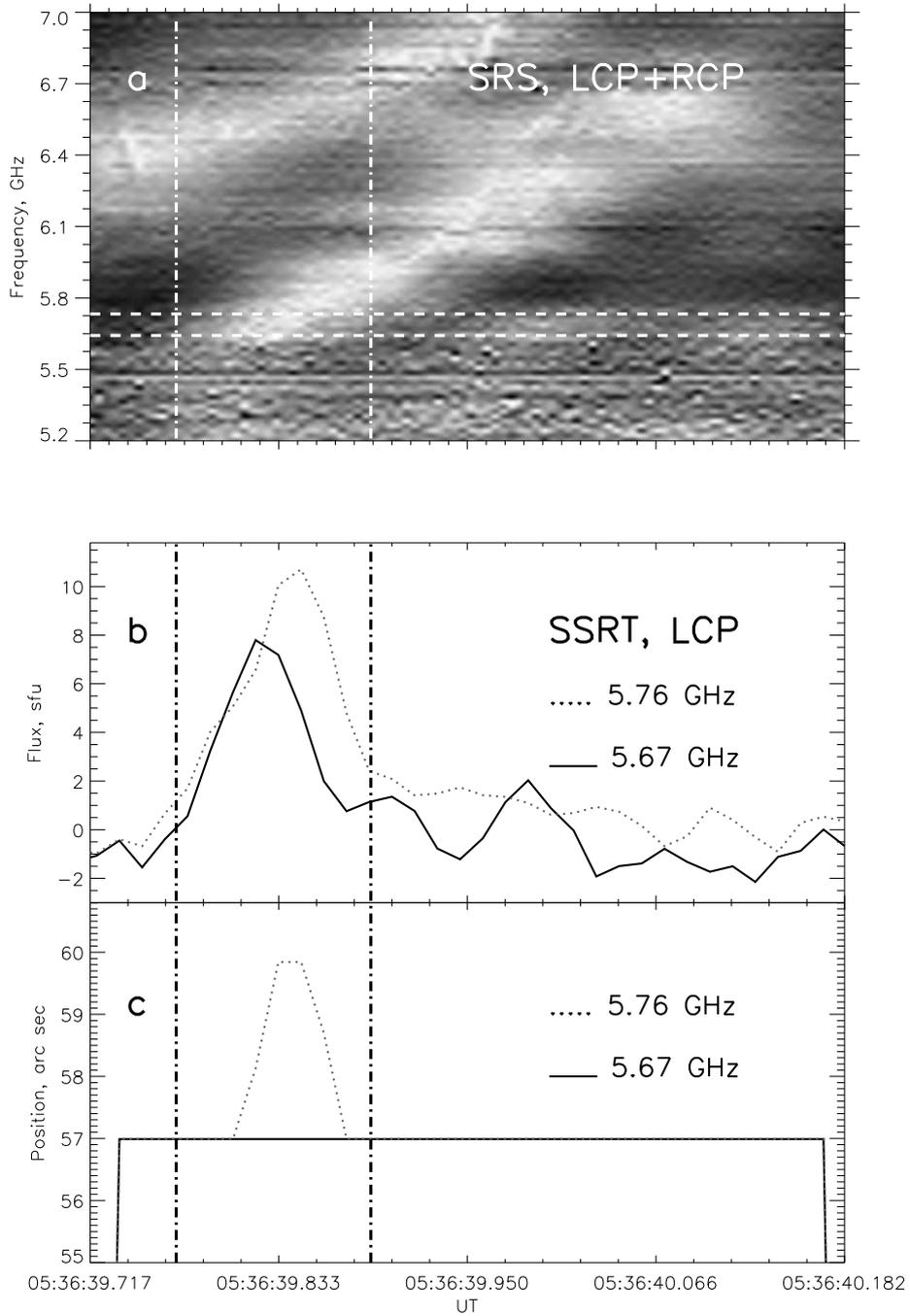}
              }
               \caption{
Observations on 14 April 2002. A subsecond pulse recorded at two
frequencies. (a) Dynamic spectrum, 5.2 - 7.0 GHz band (Stokes I,
SBRS). Horizontal dashed lines show the SSRT observing band. (b)
Time profiles at 5.76 GHz (dotted) and 5.67 GHz (solid). (c)
Positions of subsecond sources (centroids) observed at the two
SSRT frequencies vs. time. Vertical dash-dotted lines indicate the
SSP interval.
                      }
   \label{F2-simple}
   \end{figure}

   \begin{figure}
  \centerline{\includegraphics[width=\textwidth]{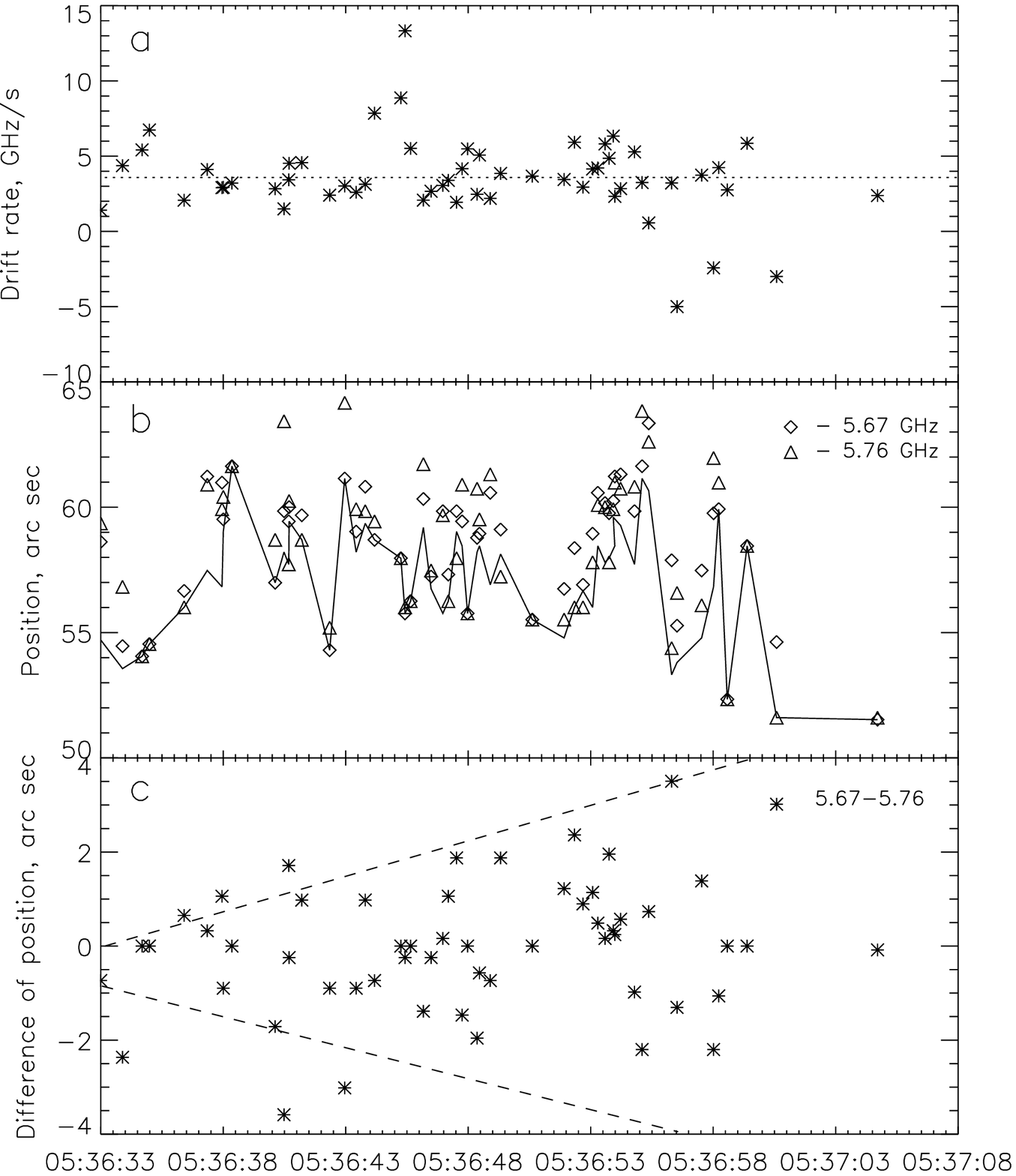}
              }
              \caption{a) Drift rates (horizontal dotted line marks the mean value);
              b)  Displacements of
impulsive sources in 1D scans at 5.67 GHz (diamonds) and at 5.76
GHz (triangles). The solid line shows centroids of the background
burst; c) Absolute values of difference between positions at two
frequencies.
                      }
   \label{F3-simple}
   \end{figure}

\section{Observations}

The 14 April 2002 impulsive flare occurred during the interval
05:36 -- 05:48 UT, with an SXR maximum of C5.3 at 05:37 UT. The
flaring active region AR9907 was located at S03E45.

Subsecond microwave structures occurred at 5.2 -- 7.6 GHz during
the one-minute interval of the flare rise phase. According to GOES
data the narrow-band structures have appeared after an abrupt
increase of temperature 4 up to 10 MK and the emission measure
rise to $8\times 10^{47} \mathrm{cm}^{-3}$
(Figure~\ref{F1-simple}a). The HXR data (RHESSI) were no available
for this period.

From the SBRS dynamic spectrum we have extracted 56 bursts, whose
flux exceed 3$\sigma$ and were recorded within the SSRT band
(Figure~\ref{F1-simple}). Comparison between RCP and LCP signals
shows that the background burst was weakly polarized at 5.7 GHz
(under to 5\%), whereas subsecond pulses were pronouncedly
left-polarized. Noteworthy is the high degree of agreement between
the time profiles from SBRS and SSRT. The drifting structures
occurred in the entire SBRS frequency range and the total spectral
width of the individual bursts was found to be up to 1.5 GHz.

Figure~\ref{F2-simple}a shows an example of a fast-drifting burst
dynamic spectrum and the SSRT time profiles at two frequencies
from 05:36:39.7 to 05:36:40.1. The horizontal dashed lines mark
the frequencies of 5.67 and 5.76 GHz, at which the NS linear
interferometer observed the burst. Figure~\ref{F2-simple}b shows
LCP emission component at two SSRT frequencies. The positions of
the impulsive source centroids are shown in
Figure~\ref{F2-simple}c at moments when the sources are observed
at the SSRT frequencies. The positions are calculated
automatically for the intervals when the rapidly drifting flux
component exceeds 10\% of the gradual background burst. In this
example the fluxes were sufficiently large in the interval marked
by the vertical lines and the centroid of its 1D image at 5.67 GHz
was displaced by about 3$^{\prime\prime}$ relative to the
background burst centroid.

The simultaneous records at two SSRT frequencies gave us a rare
opportunity to estimate the plasma density gradient and velocity
of exciter independently. For each pulse we defined the SSP
position using the localization technique for sources of temporal
fine structure \cite{Meshalkina04} as well as their displacements
and drift rates.

Measurements for the pulses are shown in Figure ~\ref{F3-simple}.
The scatter of drift rates is obviously random in time and
distributed around the average value 3.6 GHz $s^{-1}$(Figure
~\ref{F3-simple}a) except the four pulses. During the 2001 April
14 event drifting bursts were observed at the frequencies of 5.67
and 5.76 GHz ($\Delta f = 0.09$ GHz). The 1D brightness
distributions were recorded every 14 ms in a direction
perpendicular to the NS line in Figure 4 and show the spatial
characteristics along the flare loop. The position variation of
the background burst provides guidance on accuracy of absolute
spatial measurements (Figure~\ref{F3-simple}b). The positions of
the drifting burst sources were measured relative to the
co-temporal burst centroid. The spaces between the "gravity"
centers of pulse sources in 1D data recorded at different
frequencies show the projected displacement ($\Delta l$) of the
exciter. It is seen that spatial scatter of displacement $\Delta
l$ is increasing in time (dashed lines at
Figure~\ref{F3-simple}c). So, the density profile was likely to
flatten out along the loop.

\begin{figure}
  \centerline{\includegraphics[width=0.8\textwidth,clip=]{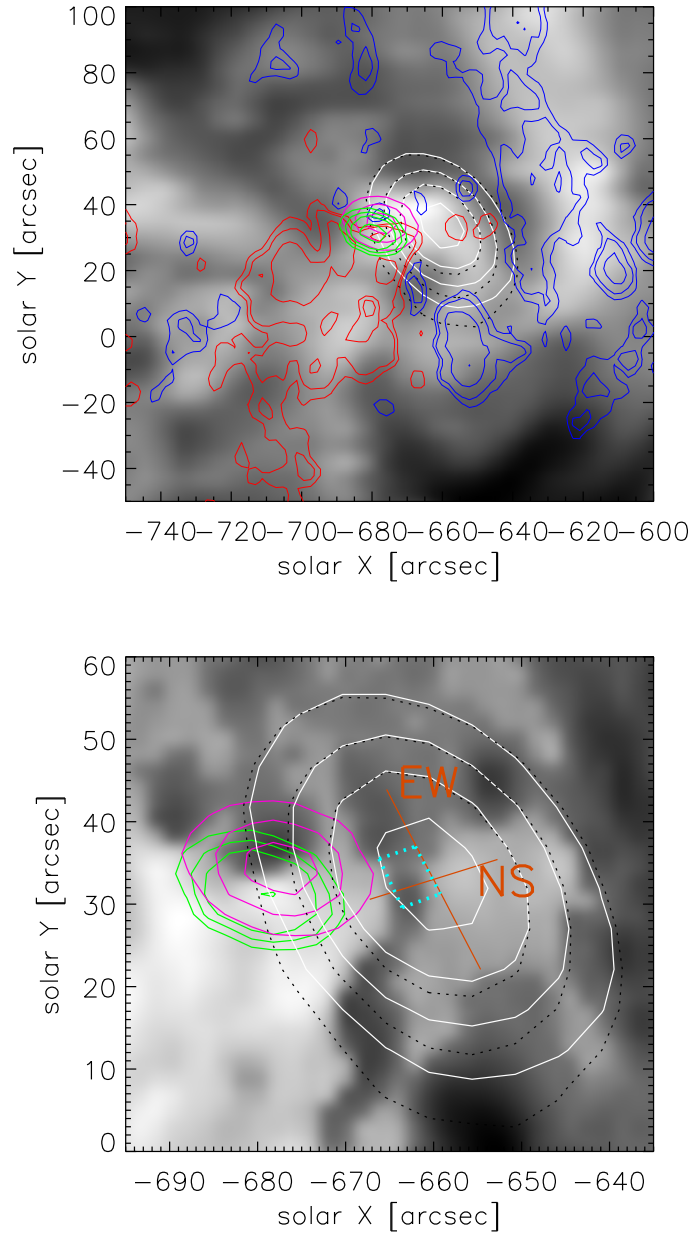}
                 }
              \caption{Top: Contours of 5.7 GHz (SSRT, difference between
  images at 05:36 and 05:34) are sources
 (white solid: Stokes I, 30\%, 50\%, 70\%, 90\% of the maximum; black dotted:
 Stokes V, LCP, 30\% , 50\%, 70\% of the maximum). Pink contours:
 Stokes I, (NoRH, 17 GHz, 05:37,  50\%, 70\%, 90\% of the maximum);
 green contours:
Stokes V(NoRH, 17 GHz , 05:37,  50\%, 60\%, 70\%, 90\% of the
minimum) superimposed on an EIT image 195 $\AA$ (05:36:05). The
axes show arc sec from the solar disk centre. Red contours
represent positive polarity of MDI magnetogram (06:24:30), blue
contours show negative polarity (levels $\pm100G, \pm200G,
\pm500G, \pm1000G, \pm1500G$). Bottom: Enlarged part of image in
panel (a). The meaning of contours and levels are the same.
Background is MDI magnetogram (06:24:30, light areas denote
positive polarity; dark areas negative polarity). The blue frame
restricts the scatter of positions of subsecond pulse sources. The
orange lines show the scan directions of the SSRT linear
interferometers.
                       }
   \label{F4-simple}
   \end{figure}

The spatial structure of the flare is shown in
Figure~\ref{F4-simple}. The microwave image at 5.7 GHz shows a
loop, lying approximately along the northwest direction. The
negatively polarized source (LCP) at 17 GHz was located near the
northern footpoint of the loop as observed in UV
(Figure~\ref{F4-simple}a). The apparent sizes of subsecond pulse
sources were below the SSRT beam width, microwave burst size was
about 25$^{\prime\prime}$.

The centroids of both the background and impulsive emissions were
located close to the loop top. Centroids of the fast drifting
sources were located within the blue dotted frame of $5\times
7^{\prime\prime}$. The distance between the loop footpoints was
about 45$^{\prime\prime}$ as seen in UV.

\begin{figure}
  \centerline{\includegraphics[width=0.8\textwidth,clip=]{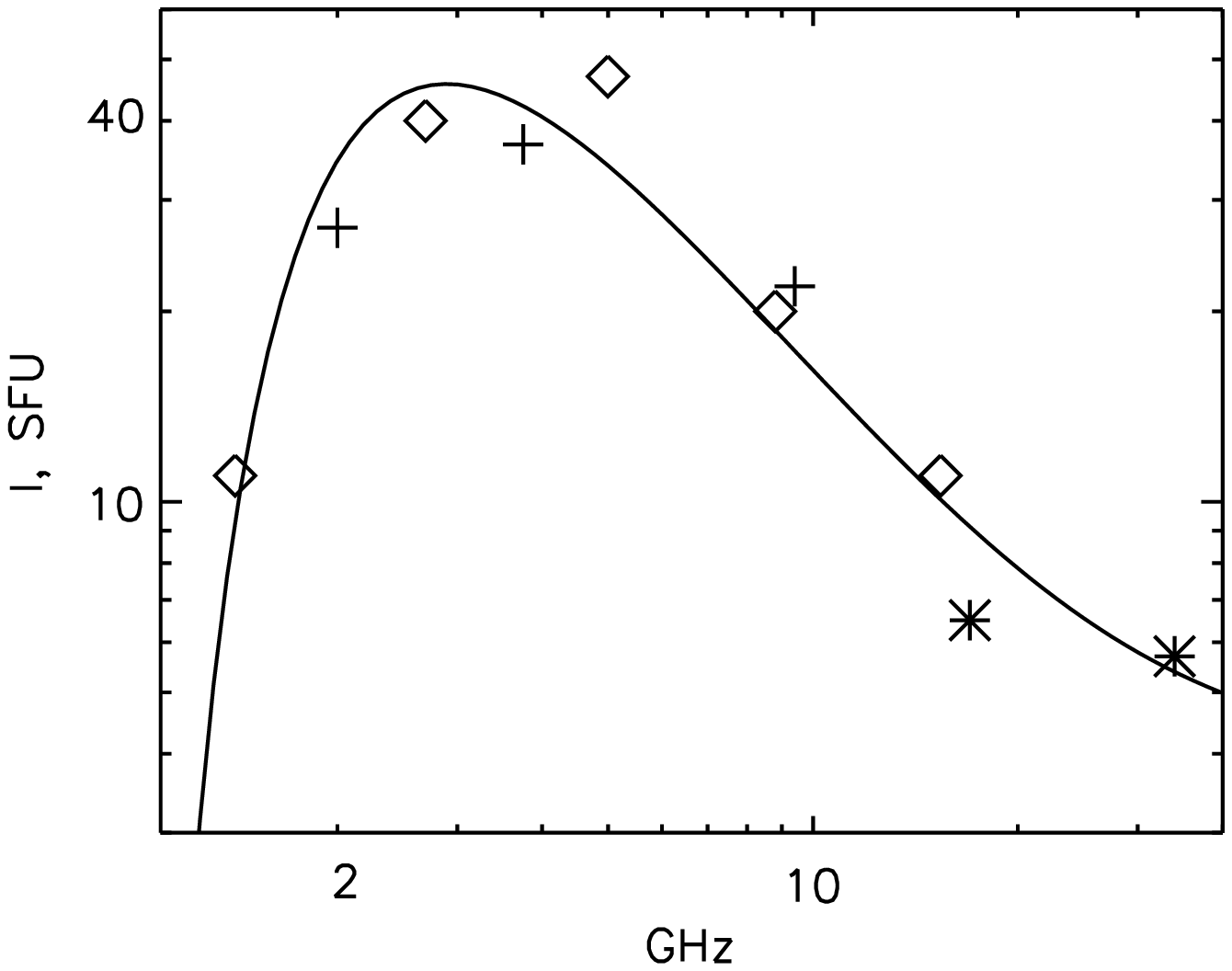}
                 }
              \caption{Spectrum of the background microwave burst
during observations of type III bursts (05:36:41.00). Crosses are
for data from NoRP spectropolarimeters; diamonds denote the
Learmonth observatory of the RSTN network; asterisks stand for
fluxes obtained by integrating brightness temperatures over
sequences of NoRH flare region images at 17 and 34 GHz. The solid
curve shows calculation results for gyrosynchrotron and
bremsstrahlung using the code by Fleishman and Kuznetsov (2010).}
   \label{F5-simple}
   \end{figure}

The spectrum of the microwave burst was calculated using the code
of Fleishman and Kuznetsov \shortcite{Kuznetsov10} and parameters
from observational data (area is $1.3\times10^{19} $cm$^{2}$,
thickness is $3.6\times10^{9}$ cm) (Figure~\ref{F5-simple}). A
satisfactory fitting can be obtained based on the following
reasonable parameters: plasma temperature 10 MK and density
$1.1\times10^{10}$ $\mathrm{cm}^{-3}$, electron spectrum exponent
4, density of nonthermal electrons with over 16 keV energies
$8\times10^6$ cm$^{-3}$, magnetic field 70 G, line of sight 80
degrees to the magnetic field direction.

\section{Discussion}

The fast drifting bursts were observed in the total SBRS range
from 5.2 up to 7.5 GHz. It is well established that the
frequencies of fast drifting burst in microwaves are close to the
local harmonic Langmuir frequency \cite{Fleishman98, Altyntsev00,
Benz00}. So the plasma density was in the range of $(1-2)\times
10^{11}$ $\mathrm{cm}^{-3}$ in the emission volume during the
total time interval with the fine spectral structures.

The emission region was located near the loop top within the
$5\times 7^{\prime\prime}$ frame. The projected displacement along
the loop (Figure~\ref{F3-simple}) were small in most events and as
a rule did not exceed the measurement accuracy of two arcseconds.
So, the longitudinal size of emission region was small in
comparison with the loop length and did not exceed three-five
thousand km. This contradicts the usual interpretation of positive
drifting bursts by exciter moving from loop top towards denser
loop foot points.

Using the emission measure EM calculated from GOES data and
assuming a cylindrical shape of emission region we can estimate
the radius value as $(1-2)\times 10^8$ cm for the emission value
of 5 $\times 10^{47}$ $\mathrm{cm}^{-3}$. Here we have taken the
cylinder length of $3\times 10^{8}$ cm and plasma density of
$(1-2)\times 10^{11}$ $\mathrm{cm}^{-3}$. It is an over-estimation
because we assume that the increase of the GOES signals was due to
the dense emission region. The calculated sizes agree with the
frame bounded in Figure~\ref{F4-simple}c. Note that the position
scattering of the fast-drifting sources was gradually rising along
the loop (Figure~\ref{F3-simple}c). So, the density profile along
the loop is flattering out at the approximately constant level and
the GOES emission measure growth was caused by the dense region
expansion.

The recorded fast-drifting bursts have started after the
temperature rising up to 10 MK. Using the Equation (1) and taking
$H= 3\times 10^{8}$ cm we estimated that this region became
optically thin ($\tau < 1$) at this time. The dense emitting
region was surrounded by a low-density plasma as follows from the
fitting of the continuum burst spectrum.

The frequency drift rate, $df/dt$, is determined by the
expression:

\begin{eqnarray}
df/dt = \frac{A}{2\sqrt{n}} \frac{dn}{dt} = \frac{A}{2\sqrt{n}}
 \left( \frac{\partial n}
{\partial t} + \frac{\partial n} {\partial l} \frac{\partial l}
{\partial t} \right)= \frac{A}{2\sqrt{n}}
 \left( \frac{\partial n}
{\partial t} + \frac{\partial n} {\partial l} v \right)
\label{eqn1},
\end{eqnarray}

\noindent where coefficient $A=\frac{\alpha}{2\pi}
\sqrt{\frac{4\pi e^2}{m}}$, $\alpha=2$ for the harmonic,
$\frac{\partial n} {\partial t}$ describes how the plasma density
changes with time, $\frac{\partial n} {\partial l}$ is an electron
density gradient along the path through which the exciter moves
with velocity $v$. The equation can be rewritten as $df/dt \approx
2 \Delta f \left( \frac{1} {\tau_\mathrm{ssp}} - \frac{v} {\Delta
l} \right)$, where ($\tau_\mathrm{ssp}$) is a temporal scale of
density change.

\begin{figure}
  \centerline{\includegraphics[width=\textwidth,clip=]{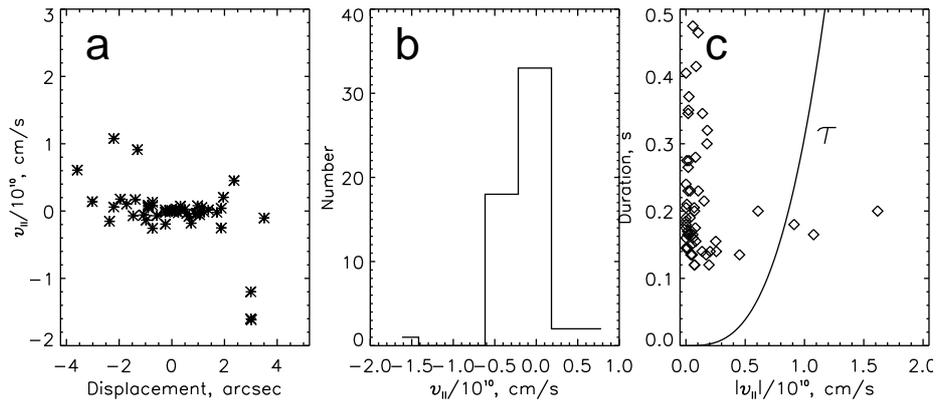}
                 }
                 \caption{(a) Exciter velocities $v$ \textit{vs.} displacements
$\Delta l$; northward motion is positive. (b) Histogram of exciter
velocities. (c) Relationship between the calculated absolute
values of the exciter velocities and the total durations of
drifting bursts. The solid curve shows calculated lifetimes
$\tau$.
                 }
                 \label{F6-simple}
   \end{figure}

Generally, only the second term $\frac{\partial n} {\partial l}
v$, is considered in the interpretations of drifting bursts. For
the frequency drift 3.6 GHz s$^{-1}$ and $\Delta l = 1.5$ Mm the
exciter velocity is rather small ($3 \times 10^7$ cm s$^{-1}$)
compared to a velocity of nonthermal emitting electrons. The
dependences in Figure~\ref{F6-simple} are calculated in the
approximation of $ 2\Delta f/{\tau_\mathrm{ssp}}$ = 3.6 GHz/s. For
density growth rate ${\partial n}/ {\partial t} \approx
0.6\times10^{11} $cm$^{-3} $s$^{-1}$ the velocity along the loop
tends to zero. Therefore, this value is the upper limit of the
density growth rate.

The exciter velocity is found to be in the range $-1.6$ to
$1.2\times10^{10}$ cm s$^{-1}$, but for most pulses the
longitudinal component of the velocity is less than $3\times 10^9$
cm s$^{-1}$. The upper estimate of the life-time is determined by
the Coulomb collisions ($\tau =3.1 \times 10^{-20} {v^3 / n}$;
\opencite{Trubnikov65}; \opencite{ Benz93}) and is shown by a
solid line in Figure~\ref{F6-simple}c. The diamonds denote
observable pulse durations. The observable durations considerably
exceed the electron life-times estimated from energies calculated
as $mv$$_{||}^{2}/2$. In this case we must assume for most pulses
that the transverse velocity of the emitting electrons is larger
than the longitudinal velocity component. So the electrons are
mainly accelerated across the magnetic field. Note that influence
of relativistic effects on the velocity measurements caused by a
velocity component along the line of sight is rather small because
of the low exciter velocity and the small cross-section size of
the emission region.

The analysis showed that the derived distributions in
Figure~\ref{F6-simple}b,c with and without taking into account the
first term of Equation (2) are very similar to ones found for 30
March 2001. In other words the most of the electrons are
definitely accelerated across magnetic field.

Usually the observation of type III-like bursts is considered as
direct evidence of electron beams propagating in the solar corona.
At large pitch-angle anisotropy of the emitting electrons and
small exciter velocities along the loop the most likely mechanism
should be loss-cone instabilities. Such distribution is able to
excite hybrid plasma waves (\opencite{Stepanov1974};
\opencite{Zaitsev83}; \opencite{Fleishman98}). Electromagnetic
emission at the double plasma frequency should be produced due to
the nonlinear interaction between upper hybrid waves. In this
paper we did not discuss the emission mechanism in detail because
there are a number publications on this subject.

The modern models based on magnetic reconnection processes assume
electron acceleration in multiple acceleration sites
(\opencite{Zharkova}; \opencite{Lin}). Rapid variations in flare
emissions imply that reconnection is non-steady and a time-varying
electric field is present in a reconnecting current sheet
(\opencite{Litvinenko}). Multiple episodes of accelerations can be
produced by repetitive interactions of multiple magnetic islands
formed in regions either at the top of single helmet-like loop or
at the intersection of interacting loops (\opencite{Drake};
\opencite{Barta}).

It is reasonable to suggest that an individual drifting burst is a
response to elementary energy release.  In compressible plasma the
pulses of electron acceleration should be accompanied by density
growths at the reconnection site (\opencite{Priest00}). This
explains the observed positive frequency drift pulses if the
acceleration and reconnection sites are coincide.

\section{Conclusions}

The flare of 14 Apr 2002 with tens of short drifting bursts
simultaneously recorded by the SSRT and wide-band
spectropolarimeters SBRS was analyzed. The SSRT have recorded the
sources of the drifting bursts simultaneously at two frequencies
(5.67 and 5.76 GHz) which allowed us to distinguish between the
spatial and temporal characteristics.  The measured drift rates
are clustered around value of 3.6 GHz s$^{-1}$. It was shown that
the sources with emitting electrons are located inside a compact
dense region near the loop top. The sizes of the acceleration
region does not exceed a few Mm. Thus, observations and analyses
of short-duration microwave drifting bursts provide a meaningful
method to obtain information on the magnetic reconnection process.
We hope that future spectrometers operating in a wide frequency
range will provide us with a greater number of events for
research.

\section*{Acknowlegement}
We are grateful to the teams of Nobeyama Radio Observatory, RSTN,
who have provided free access to their data. The authors thank the
anonymous referee for useful suggestions. The research carried out
by Robert Sych at NAOC was supported by the Chinese Academy of
Sciences Visiting Professorship for Senior International
Scientists, grant No. 2010T2J24. The research by Yihua Yan and
Chengming Tan was supported by NSFC and MOST grants (10921303,
2011CB811401). This study was supported by the Russian Foundation
of Basic Research (12-02-91161, 12-02-00173, 12-02-10006) and by a
Marie Curie International Research Staff Exchange Scheme
Fellowship within the 7th European Community Framework Programme.
The work is supported in part by the grants of Ministry of
education and science of the Russian Federation (State Contracts
16.518.11.7065 and 02.740.11.0576).

\end{article}
\end{document}